\documentclass[notitlepage,11pt]{revtex4-2}

\usepackage{graphicx}
\usepackage{dcolumn}
\usepackage{bm}
\usepackage[latin1]{inputenc}
\usepackage{epstopdf}
\usepackage{amsmath}
\usepackage{upgreek}  
\usepackage{amssymb}
\usepackage{esint}
\usepackage{array}
\usepackage{xcolor}
\usepackage{latexsym}
\usepackage{mathrsfs}
\usepackage[sans]{dsfont}
\usepackage[colorlinks=true,breaklinks=true,allcolors=blue]{hyperref}

\begin{document}

\preprint{}

\title{Miscibility regimes in a $^{23}$Na-$^{39}$K quantum mixture}

\author{E.M. Gutierrez}
\affiliation{Instituto de F\'{\i}sica de S\~{a}o Carlos, Universidade de S\~{a}o Paulo, C.P. 369, 13560-970 S\~{a}o Carlos, SP, Brazil}
\author{G.A. de Oliveira}
\affiliation{Instituto de F\'{\i}sica de S\~{a}o Carlos, Universidade de S\~{a}o Paulo, C.P. 369, 13560-970 S\~{a}o Carlos, SP, Brazil}
\author{K.M. Farias}
\affiliation{Instituto de F\'{\i}sica de S\~{a}o Carlos, Universidade de S\~{a}o Paulo, C.P. 369, 13560-970 S\~{a}o Carlos, SP, Brazil}
\author{V.S. Bagnato}
\affiliation{Instituto de F\'{\i}sica de S\~{a}o Carlos, Universidade de S\~{a}o Paulo, C.P. 369, 13560-970 S\~{a}o Carlos, SP, Brazil}
\author{P.C.M. Castilho}
\email{patricia.castilho@ifsc.usp.br}
\affiliation{Instituto de F\'{\i}sica de S\~{a}o Carlos, Universidade de S\~{a}o Paulo, C.P. 369, 13560-970 S\~{a}o Carlos, SP, Brazil}

\begin{abstract}
Effects of miscibility in interacting two-component classical fluids are relevant in a broad range of daily applications. When considering quantum systems, two-component Bose-Einstein condensates provides a well controlled platform where the miscible-immiscible phase transition can be completely characterized. In homogeneous systems, this phase transition is governed only by the competition between intra- and inter-species interactions. However in more conventional experiments dealing with trapped gases, the pressure of the confinement increases the role of the kinetic energy and makes the system more miscible. In the most general case, the miscibility phase diagram of unbalanced mixtures of different atomic species is strongly modified by the atom number ratio and the different gravitational sags. Here, we numerically investigate the ground-state of a $^{23}$Na-$^{39}$K quantum mixture for different interaction strengths and atom number ratios considering realistic experimental parameters. Defining the spatial overlap between the resulting atomic clouds, we construct the phase diagram of the miscibility transition which could be directly measured in real experiments.
\end{abstract}

\maketitle

\section{Introduction}

Mixtures of quantum fluids such as superfluid $^{3}$He-$^{4}$He~\cite{Reppy1967PhaseSepHelium,Treiner1995PhaseSepHelium,Dietrich2004PhaseDiagramHelium,bennemann1976physics} and atomic Bose-Einstein condensates (BECs)~\cite{stenger1998spin,PhysRevLett.78.586,PhysRevLett.81.1539,PhysRevLett.97.120403,Papp2008TunableBEC,PhysRevA.84.011603,PhysRevA.88.023601,wang2015double,PhysRevA.92.053602,Torben2018FeshbachNaK} exhibit different miscibility regimes as a result of the competition between intra- and interspecies interactions between its components. The high level of control of the latter (mass and atom number ratio between the atomic components, temperature, interaction strengths, confinement and system dimensionality) had allow the observation of a large variety of physical phenomena not accessible with single component systems. In optical lattices, new phase transitions gives rise to a much more complex phase diagram than the simple extension of the superfluid to Mott insulator transition~\cite{altman2003phase,isacsson2005superfluid}; polaron physics can be explored with large imbalanced mixtures~\cite{bruderer2008self,spethmann2012dynamics}; and the recently observed self-stabilized quantum droplets with liquid-like behaviour can be produced when beyond mean-field effects became dominant~\cite{cabrera2018quantum,semeghini2018self,PhysRevResearch.1.033155,guo2021lee}. The miscibility regime of the system plays a fundamental role on the superfluid properties of the mixture directly affecting the observation of the mentioned new phenomena.

As for its classical counterpart, a mixture of two fluids is miscible if the fluids totally overlap forming a homogeneous solution or immiscible if the fluids remain phase-separated~\cite{khabibullaev2003phase,dagotto2013nanoscale}. In the case of homogeneous quantum fluids, the miscible-immiscible phase transition is well defined and it is mediated by the miscibility parameter~\cite{pethick2008bose,pitaeviskii2003BECbook}
\begin{equation}\label{Eq:delta}
    \delta = \frac{u_{12}^{2}}{u_{11}u_{22}}-1,
\end{equation}
where $u_{11}$ and $u_{22}$ are the intraspecies interaction coupling constants of species 1 and 2, respectively, and $u_{12}$ gives the interspecies interaction. This is an intuitive parameter based on the competition between intra- and interspecies interactions: if $u_{12}$ overcomes the intraspecies interaction terms ($\delta>0$), the fluids strongly repel each other making the system immiscible. On the contrary, if $u_{12}$ is smaller than the intraspecies interactions ($\delta<0$), the fluids overlap and the system is miscible. In such a picture, the miscibility regime of a two-component quantum gas can be controlled by varying the interaction coupling terms, which can be experimentally realized with the use of Feshbach resonances~\cite{Chin2010Feshbach}.

However, until very recent~\cite{Gaunt2013HomogeneousBEC}, homogeneous atomic BECs were not experimentally produced. Instead, trapped atomic BECs exhibit an inhomogeneous density distribution as a result of the confinement. The increased role of the kinetic energy in such systems contributes to a more miscible mixture where phase-separation occurs for larger $u_{12}$ than the condition set by Equation~\ref{Eq:delta}. The shift at the miscible-immiscible critical point has been obtained in the case of mixtures composed of distinct hyperfine states of the same atomic species~\cite{Wen2012PhaseSepAndConfinement,Navarro2009PhaseSep,Bisset2018QuantumSpinFluct2BECs}. In the more broad scenario of unbalanced mixtures of different atomic species, the atom number ratio~\cite{Lin2020AtomNumberMiscibility}, the mass imbalance and the difference in trapping configurations between the components were also shown to affects the boundary of the miscibility phase transition~\cite{PhysRevA.62.053601,PhysRevA.70.063606,Arlt2016PhaseSeparation,Cikojevi__2018}. The contribution of gravity, relevant for all real experiments due to the induced gravitational sag~\cite{PhysRevLett.112.250404,PhysRevLett.117.145301}, is rarely taken into account in numerical simulations. 

In this work, we perform numerical simulations of the ground-state of a two-component quantum mixture of $^{23}$Na and $^{39}$K atoms for different interaction strengths, according to the relevant Feshbach resonances for magnetic fields in the range of $95-117$~G~\cite{Chiara2007fesh39K,Torben2018FeshbachNaK}, in order to show the realistic miscibility regimes accessible in the experimental setup being developed in our laboratory~\cite{castilho2019NaK} in the presence of gravity. We explore the effect of changing the number of atoms of the minority species ($^{39}$K), therefore changing the atom number ratio $\eta$, and calculate the spatial overlap between the atomic clouds as a quantity able to characterize the change in the miscibility regime of the system. The numerical simulations are performed at zero temperature, which satisfactory reproduces the experimental results for the case of strongly degenerate atomic mixtures~\cite{PhysRevA.98.063616}, although theoretical works at finite temperature have shown a change of the miscibility condition of the system favoring phase separation~\cite{PhysRevLett.123.075301,PhysRevA.102.063303,roy2021finite}.

The article is organized as follows. In Section~\ref{Sec:Methods}, we describe the two-component quantum gas at zero temperature in terms of a pair of coupled Gross-Pitaevskii equations (GPEs)~(\ref{subsec:GPEs}) and the numerical simulation method used to obtain the ground-state of the system~(\ref{SubSec:Simulation}). In Section~\ref{Sec:Results}, we first present our experimental system producing the $^{23}$Na-$^{39}$K atomic mixtures~(\ref{SubSec:ExpSetup}), followed by the results of the numerical simulation performed with realistic experimental parameters~(\ref{subsec:ground-state}) and the construction of the phase diagram of the miscible-immiscible transition for such a mixture~(\ref{subsec:phaseDiagram}). Finally, in Section~\ref{Sec:discussion}, we highlight our main findings and discuss some future perspectives for identifying the miscibility regime of a quantum mixture comparing with the results presented in this article.

\section{Methods}\label{Sec:Methods}

\subsection{Description of an atomic quantum mixture}\label{subsec:GPEs}

Consider a mixture of two different bosonic atoms, labeled $1$ and $2$, at $T=0$ in the weakly interacting regime where interactions are treated as contact interactions. Let $N_{1}$ and $N_{2}$ be the number of particles and $\phi_{1}(\bold{r})$ and $\phi_{2}(\bold{r})$ be the corresponding normalized single-particle wave functions. In such a picture, and neglecting terms of the order of $1/N_{1}$ and $1/N_{2}$, the energy functional of the system~\cite{griffin1996bose,pethick2008bose,pitaeviskii2003BECbook} can be written as
\begin{equation}\label{Eq:EnergyFunc}
\begin{split}
E= & \int d\textbf{r}\left[\frac{\hbar^{2}}{2m_{1}}|\nabla\psi_{1}|^{2}+\vartheta_{1}(\textbf{r})|\psi_{1}|^{2}+\frac{\hbar^{2}}{2m_{2}}|\nabla\psi_{2}|^{2}+\vartheta_{2}(\textbf{r})|\psi_{2}|^{2}\right. \\
& \left.+\frac{1}{2}u_{11}|\psi_{1}|^{4}+\frac{1}{2}u_{22}|\psi_{2}|^{4}+u_{12}|\psi_{1}|^{2}|\psi_{2}|^{2}\right],
\end{split}
\end{equation}
where $m_i$ (with $i=1,2$) is the mass of atomic species $i$, $\vartheta_i(\bold{r})$ is the corresponding external potential, $u_{ii}=4\pi\hbar^2a_{ii}/m_{i}$ are the intra-species interaction terms and $u_{12}=2\pi\hbar^2a_{12}/m_{12}$ is the inter-species interaction term with $m_{12}=m_1m_2/(m_1+m_2)$, the reduced mass of the system. For all relations, $a_{ij}$ is the associated two-body $s$-wave scattering length. The wave-functions $\psi_1(\bold{r})$ and $\psi_2(\bold{r})$ are the condensate wave-function of each atomic species, defined as
\begin{equation}
    \psi_1(\bold{r})=\sqrt{N_1}\phi_1(\bold{r}) ~~\text{and} ~~ \psi_2(\bold{r})=\sqrt{N_2}\phi_2(\bold{r}).
\end{equation}

Minimizing the energy functional of Equation~\ref{Eq:EnergyFunc} under the constraint of fixed number of particles, $N_1$ and $N_2$, one obtains the time-independent coupled Gross-Pitaevskii equations
\begin{equation}\label{eq:TwoBECTimeIndGPE1}
\left[-\frac{\hbar^2}{2m_{1}}\nabla^2+\vartheta_{1}(\textbf{r})+u_{11}|\psi_{1} |^{2}+u_{12}|\psi_{2} |^{2}\right]\psi_{1}=\mu_{1}\psi_{1}
\end{equation}
\begin{equation}\label{eq:TwoBECTimeIndGPE2}
\left[-\frac{\hbar^2}{2m_{2}}\nabla^2+\vartheta_{2}(\textbf{r})+u_{22}|\psi_{2} |^{2}+u_{12}|\psi_{1} |^{2}\right]\psi_{2}=\mu_{2}\psi_{2},
\end{equation}
where $\mu_1$ and $\mu_2$ are the chemical potential of atomic species $1$ and $2$, respectively. If the interspecies interaction vanishes ($u_{12}=0$), Equations~\ref{eq:TwoBECTimeIndGPE1} and \ref{eq:TwoBECTimeIndGPE2} are no longer coupled and each species behave as a single species atomic cloud. In this case, approximations such as the Thomas-Fermi approximation~~\cite{griffin1996bose,pethick2008bose,pitaeviskii2003BECbook}, for which the kinetic term of the GPE is neglected, can be used to find a solution for the ground-state of the system. On the other hand, when $u_{12}\neq 0$, the competition between inter- and intraspecies interactions gives rise to a phase transition from a miscible to an immiscible (phase-separated) phase when increasing the positive inter-species interaction strength. The existence of overlapping and non-overlapping regions between the atomic clouds dramatically changes the ground-state configuration of the system and it is not always possible to find analytical solutions for it, even relying on approximations~\cite{riboli2002topology}. A more powerful technique to obtain the ground-state of a trapped two-component BEC makes use of a numerical simulation with imaginary time evolution of the coupled GPEs.

\subsection{Numerical simulation of the ground-state}\label{SubSec:Simulation}

The numerical simulation used to obtain the ground-state of the two-species BEC consists of projecting onto the minimum of the GPEs each initial trial state by propagating them in imaginary time~\cite{dalfovo1996bosons}. To describe the method, let us first consider a system described by a Hamiltonian $H$ for which the time evolution of one of its eigenstates, $\psi_n(\textbf{r},t)$ with $H\psi_n(\textbf{r},0) = E_n\psi_n(\textbf{r},0)$, is easily obtained as:
\begin{equation}
    \psi_n(\textbf{r},t)=\psi_n(\textbf{r},0)e^{-i\frac{E_n}{\hbar}t},
\end{equation}
where $E_n$ is the energy associated with the $n$-eigenstate. The time evolution of an arbitrary trial function $\Psi(\textbf{r},0)$, written as a linear combination of the system's eigenstates, is simply given by
\begin{equation}\label{Eq:PsiTimeEvol}
    \Psi(\textbf{r},t)=\sum_n\psi(\textbf{r},t)=\sum_n\psi(\textbf{r},0)e^{-i\frac{E_n}{\hbar}t}.
\end{equation}
If ones calculates $\Psi(\textbf{r},t)$ for $t=-i\tau$, the complex exponentials in Eq.~\ref{Eq:PsiTimeEvol} are replaced by exponential decays with decay constants given by $E_n/\hbar$. By evaluating $\Psi(\textbf{r},t)$ at different time steps $\Delta \tau$ with $\tau=\xi\Delta \tau$, $\Psi(\textbf{r},\tau)\rightarrow\psi_0(\textbf{r},\tau)$, the ground-state of the system. The exact convergence is only obtained when $\tau\rightarrow\infty$, however, convergence methods based on the variation of the total energy of the system are used to set an upper limit for $\tau$.

In the numerical simulations performed in this work, we define a trial function $\Psi_i(\textbf{r},0)$ for each species $i$ with time evolution given by
\begin{equation}
    \Psi_{i}(\textbf{r},t)=e^{-i\frac{\widehat{H}_{i}}{\hbar}t}\Psi_{i}(\textbf{r},0),
\end{equation}
where $\widehat{H}_{i}\psi_i(\textbf{r})=\mu_i\psi_i(\textbf{r})$ from Equations~\ref{eq:TwoBECTimeIndGPE1} and~\ref{eq:TwoBECTimeIndGPE2}. Considering $t = -i\Delta\tau$ with $\Delta \tau$ infinitesimal, the resulting exponential can be expanded in a Taylor series and the time evolution of $\Psi_{i}(\bold{r},t+\Delta \tau)$ is given by
\begin{equation}\label{Eq:Sim}
    \Psi_{i}(\textbf{r},t+\Delta \tau)\approx \Psi_{i}(\textbf{r},t)-\frac{\widehat{H}_{i}}{\hbar}\Psi_{i}(\textbf{r},t)\Delta \tau.
\end{equation}
In order to achieve sufficient long times in the simulations let it be $t_{\text{final}}=\xi\Delta t$, with $\xi$ being an integer, Equation~\ref{Eq:Sim} is calculated $\xi$ times. The resulting wave-function obtained after each time step is normalized in order to preserve the atom number.

\section{Results}\label{Sec:Results}

The numerical simulations performed in this work are done following the parameters of the experimental setup being developed in our laboratory. For this reason, we first start the Results Section, subsection~\ref{SubSec:ExpSetup}, with a description of the experimental setup and its current status in the preparation of a two-species BEC of $^{23}$Na and $^{39}$K. Later, the results from the numerical simulations are presented and discussed in the following two subsections.

\subsection{Experimental setup}\label{SubSec:ExpSetup}

A complete description of the experimental setup and experimental sequence for producing a Bose-Einstein condensate of $^{23}$Na atoms is described in~\cite{castilho2019NaK}. Here, we present a short description of the system giving the experimental parameters relevant for the simulations performed later in this Section.

Briefly, sodium and potassium atoms coming from independent two-dimensional magneto-optical traps (2D-MOTs)~\cite{Catani2006MOTK,lamporesi2013MOTNa} are combined in a common vacuum chamber where they will be trapped and further cooled in a three-dimensional MOT (3D-MOT). Due to the strong interspecies losses present in the Na-K mixture~\cite{wu2013strongly,castilho2019NaK}, the operation of an intial two-color MOT is not the best alternative in our experiment. Instead, we chose to favor the minority species (potassium) during the MOT phase, starting the MOT sequence with the loading of a single species MOT of $^{39}$K until it reaches the saturation value ($\sim 20~$s). Next, we operate the two-color MOT by switching on the lights responsible for trapping and cooling sodium atoms. We control the initial atom number ratio $N_{\text{Na}}^0/N_{\text{K}}^0$ by changing the time duration of the loading of the sodium atoms in the two-color MOT operation.

Once the two species are loaded, we perform subsequent cooling procedures followed by a fine pumping stage which transfer both species to the $F=1$ ground-state before turning on an optically plugged Quadrupole trap~\cite{Davis1995BECNa}. At the beginning of the magnetic trap the atomic clouds have $N_{\text{Na}}\sim 1\times 10^9~$atoms and $N_{\text{K}}\sim 1\times 10^6~$atoms both at $T = 220~\mu$K trapped in the $\vert F=1, m_{F}=-1>$ hyperfine ground-state.

Evaporative cooling~\cite{Evap1988} of sodium is done with microwave radiation at $\sim 1.7~$GHz while potassium atoms are sympathetic cooled~\cite{Schreck2001sympathetic,modugno2001BECK} decreasing its temperature without significant atom loss. At $T\sim 6-7~\mu$K, the atomic clouds are transferred to a pure optical dipole trap (ODT)~\cite{grimm2000odt} where the interspecies interaction can be tuned with the use of Feshbach resonances~\cite{Chin2010Feshbach} by applying a uniform magnetic field. We have atomic clouds with $N_{\text{Na}} = 5\times 10^6$ and $N_{\text{K}} = 8\times 10^5$ at the beginning of the ODT for maximum atoms number of $^{39}$K. In single-species operation for sodium under the same conditions we obtain an almost pure BEC (with BEC fraction $> 80 \%$) with $N = 1\times 10^6~$atoms at $T\sim 80~\mu$K after applying an optical evaporation which reduces the initial ODT potential height by a factor of five in $4.2$~s. The final ODT configuration exhibit a planar geometry with equal frequencies in the $xy$-plane perpendicular to the gravity direction. The final frequencies are $\omega_{\text{x,y}} = 2\pi \times 107(137)~$Hz and $\omega_{\text{z}}=2\pi\times 148(193)~$Hz for Na(K), respectively. This is the actual situation of our experimental system and, following the initial atom number difference in the ODT, we estimate to be able to obtain a two-species BEC once implemented the Feshbach field. Following these experimental numbers we performed the simulations described in Section~\ref{SubSec:Simulation} which results are presented in the following.

\begin{figure}
\includegraphics[width=13.5 cm]{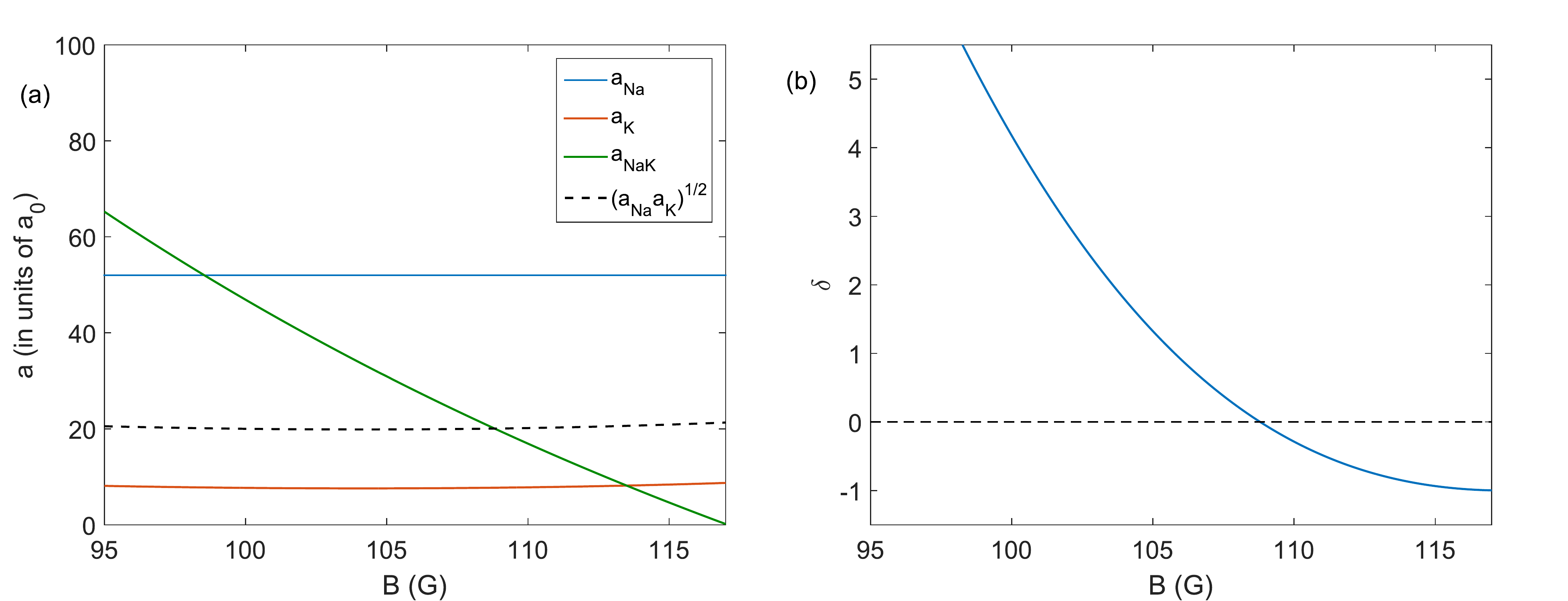}
\caption{(a) Scattering lengths as a function of the magnetic field for the intra-species interactions of $^{23}$Na, $a_{\text{Na}}$ (in blue), and $^{39}$K, $a_{\text{K}}$ (in red), and for the inter-species interaction $a_{\text{NaK}}$ (in green) considering both atoms in the $\vert F=1, m_F=-1\rangle$ hyperfine state. The black dashed line given by $(a_{\text{Na}}a_{\text{K}})^{1/2}$ represents the value of $a_{\text{NaK}}$ for which the system changes from immiscible to miscible with $\delta=0$. (b)~Miscibility parameter as a function of the magnetic field. At $B=109.1~$G with $\delta=0$ the system changes from immiscible to miscible when increasing $B$.\label{fig:Feshbach}}
\end{figure}

\subsection{Ground-state of~ $^{23}$Na~-~$^{39}$K mixtures}\label{subsec:ground-state}

The ground-state of ~$^{23}$Na~-~$^{39}$K mixtures was obtained with the numerical simulation method described in~\ref{SubSec:Simulation}. In the simulations, we discretize the space with a three-dimensional grid of $69\times 69\times 69$. The grid step size was chosen to be equal to $0.6~\mu$m resulting in a total volume of $41\times 41\times 41~\mu$m$^3$. The time interval for the simulations were $\Delta t = 50\times 10^{-6}$ in units of $1/\overline{\omega}_{1}$, where $\overline{\omega}_1=(\omega_{x}\omega_{y1}\omega_{z1})^{1/3}$ is the geometric mean of the trapping frequencies for species 1. We considered species 1 (2) as the potassium (sodium) atoms. We apply convergence methods based on the difference between the wave-functions of subsequent time intervals and monitor the total energy evolution in order to ensure the achievement of the ground-state configuration for both species. With these methods, typical integration times gave $t_{\text{final}}\sim 3000$.

The number of sodium atoms was chosen $N_{\text{Na}} = 5\times 10^5~$atoms in agreement with the numbers obtained in the experiment. The number of potassium atoms was varied with $N_{\text{K}} = 1\times 10^4 - 5\times 10^5~$atoms setting $\eta=N_{\text{Na}}/N_{\text{K}}=50 - 1$. The trapping frequencies were also set from the experimental values with $\omega_{\text{x,y}} = 2\pi \times 107(137)~$Hz and $f_\text{z}=2\pi \times 148(193)~$Hz for Na(K), respectively. The sodium scattering length was fixed to $a_{\text{Na}}=52~a_0$, with $a_0$ being the Born radius, while the scattering length of $^{39}$K, $a_{\text{K}}$, and the interspecies scattering length, $a_\text{NaK}$, was varied according to the Feshbach resonances occuring at magnetic fields smaller than $300~$G~\cite{Chiara2007fesh39K,Torben2018FeshbachNaK}. In Figure~\ref{fig:Feshbach}, we show the values of the scattering lengths ($a_{\text{Na}}$, $a_{\text{K}}$ and $a_\text{NaK}$) as a function of the magnetic field in the region with $B = 95-117.2~$G. In this region, both $a_{\text{K}}$ and $a_{\text{NaK}}$ are positive and the system changes its behaviour from immiscible to miscible with increasing the magnetic field. The predicted phase transition point for a homogeneous system (with $\delta = 0$) occurs at $B_0 = 109.1~$G~\cite{Torben2018FeshbachNaK}. The potassium scattering length was obtained with the simple relation:
\begin{equation}
    a(B) = a_\text{bg}\left(1-\frac{\Delta_1}{(B-B_{01})}-\frac{\Delta_2}{(B-B_{02})}\right),
\end{equation}
where $a_{\text{bg}}=-19a_0$ is the background scattering length, $B_{01}=32.6~$G and $B_{02}=162.8~$G are the position of the first and second resonances for $^{39}$K at the $\vert F=1, m_F=-1\rangle$ hyperfine state and $\Delta_1=55~$G and $\Delta_2=-37~$G are the corresponding resonance widths~\cite{Chiara2007fesh39K}. The $a_{\text{NaK}}$ curve displayed in Figure~\ref{fig:Feshbach} was obtained from~\cite{Torben2018FeshbachNaK} via a private communication.

\begin{figure}
\includegraphics[width=13.5 cm]{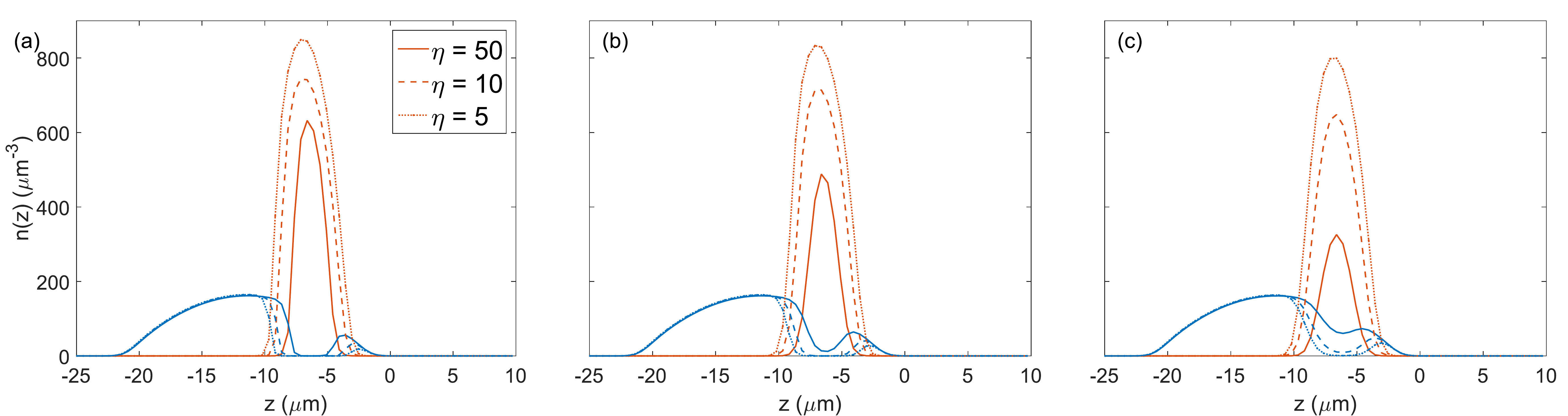}
\caption{Density profiles along the $z$-axis of the simulated ground-state of $^{23}$Na (in blue) and $^{39}$K (in red): (\textbf{a})~$B=100~$G with $\delta=4.32$, (\textbf{b})~$B=108~$G with $\delta = 0.22$ and (\textbf{c})~$B=111~$G with $\delta = -0.50$. In each case, we display the results of three atom number ratio $\eta = N_{\text{Na}}/N_{\text{K}}$ equal to 50 (solid lines), 10 (dashed lines) and 5 (dotted lines).\label{fig1}}
\end{figure}

Due to the presence of the gravitational force, each species suffers a different gravitation sag and the phase-separation at the immiscible phase occurs along the vertical direction ($z$-axis). In Figure~\ref{fig1}, we show the simulated density profiles along the $z$-axis of $^{23}$Na (in blue) and $^{39}$K (in red) for Feshbach fields $B=100~$G in~(\textbf{a}) with $\delta = 4.32$, $B=108~$G in~(\textbf{b}) with $\delta = 0.22$ and $B=111~$G in~(\textbf{c}) with $\delta = -0.50$ and different atom number ratio $\eta=50,10,5$ in solid, dashed and dotted lines, respectively.

For a fixed miscibility parameter we observe different behaviours of the system when changing $\eta$. In~Figure~\ref{fig1}~(\textbf{a}), the system is always immiscible, i.e., the sodium and potassium atoms do not share the same position in the trap for any value of $\eta$. In~Figure~\ref{fig1}~(\textbf{b}), the system is expected to be immiscible according to the miscibility parameter ($\delta=0.54>0$), however for $\eta = 50$ the phase-separated region disappears and the potassium atoms always share its position in the trap with sodium atoms, which is characteristic of a miscible system. Finally, in~Figure~\ref{fig1}~(\textbf{c}), the system is expected to be miscible with $\delta = -0.30 < 0$ but in the case of $\eta = 5$ there is still a region of the potassium cloud that do not shares the trap with sodium atoms remaining immiscible. We see that, for inhomogeneous systems, the atom number ratio has a strong influence in the miscibility regime of a two-species Bose-Einstein condensate. While in the homogeneous case, the miscibility parameter is enough to set the regime of the system, in most real experiments where the condensates are confined by a harmonic trap, additional information is necessary to establish the critical point for the miscible-immiscible phase transition.

\begin{center}
\begin{figure}
\includegraphics[width=10.5 cm]{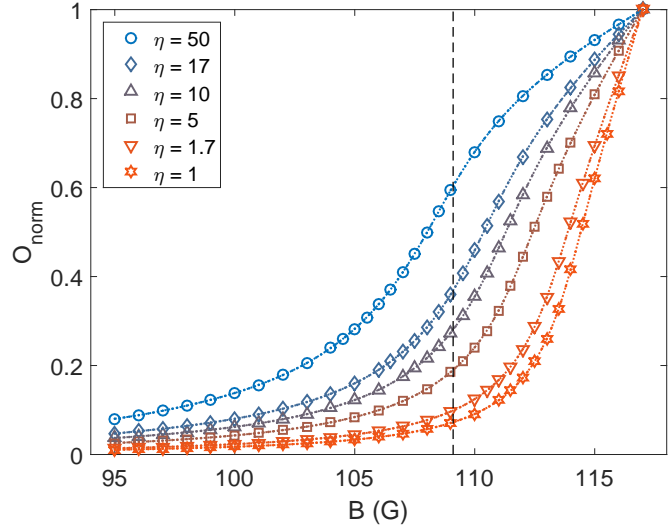}
\caption{Normalized overlap as a function of the Feshbach field for different values of $\eta$. For large $\eta$, $N_\text{K}<<N_\text{Na}$ (blue circles), the spatial overlap increases at earlier magnetic fields showing the transition to the miscible phase for $u_{12}^2>u_{11}u_{22}$ (with $\delta>0$). In the opposite scenario, for $\eta = 1$, $N_\text{K}=N_\text{Na}$ (red stars), the normalized spatial overlap significantly increases only for $B > 111~$G remaining immiscible even if $u_{12}^2<u_{11}u_{22}$ (with $\delta < 0$).\label{fig2}}
\end{figure}
\end{center}

\subsection{The miscibility phase diagram}\label{subsec:phaseDiagram}

The ground-state configurations obtained in the previous section show the flexibility of the $^{23}$Na~-~$^{39}$K mixture in achieving different miscibility regimes with the change of the Feshbach magnetic field $B$ and the atom number ratio $\eta$.

The construction of a phase diagram miscible-immiscible needs a more quantitative way of defining the miscibility region of a given set of parameters for the two-species system. Proposals to characterize the regime of such system include the calculation of the binder cumulant of the system's magnetization~\cite{PhysRevA.90.023630}, the difference between the centers of mass of each atomic cloud~\cite{Lin2020AtomNumberMiscibility}, the study of the entropy of the mixture as defined in~\cite{roy2021finite,richaud2019mixing} and the monitor of dipole oscillations of the atomic clouds in a harmonic trap~\cite{Arlt2016PhaseSeparation}. Here, similar to the works presented in~\cite{Wen2012PhaseSepAndConfinement,kumar2017miscibility}, we follow the definition of the miscible and immiscible phases and propose the calculation of the spatial overlap between the atomic clouds to be an indicator of the phase transition.

We define the spatial overlap of the atomic clouds as:
\begin{equation}
    O = \int_{-\infty}^{\infty}\vert\psi_1(\bold{r})\vert^{2}\vert\psi_2(\bold{r})\vert^{2}\text{d}\bold{r} = \int_{-\infty}^{\infty}n_1(\bold{r})n_2(\bold{r})\text{d}\bold{r},
\end{equation}
where $n_1(\bold{r})$ and $n_2(\bold{r})$ are the atomic densities of species $1$ and $2$, respectively.

In Fig.~\ref{fig2}, we show the spatial overlap normalized by the overlap, $O_0 = O(a_{12}=0)$, at each case as a function of the Feshbach field for different atom number ratio, $\eta$. The normalization was done considering that the case of vanishing inter-species interaction exhibit the maximum spatial overlap between the atomic clouds possible in each configuration. In the immiscible region ($B < B_0 = 109.1~$G), $O_\text{norm}$ exhibit small values with $O_\text{norm}\rightarrow 0$ when reducing the magnetic field for all $\eta$. Approaching $B_0$, the spatial overlap increases differently for each $\eta$ with larger $\eta$ showing an earlier increase on the spatial overlap. The limit case of $\eta=50$ shows a significant increase of $O_\text{norm}$ for $B > 104~$G, increasing over a broad range of magnetic fields. This increase occurs before the critical point estimated with the miscibility parameter, in accordance with was already observed in the density profiles of Fig.~\ref{fig1}. Reducing $\eta$ increases the magnetic field for which the spatial overlap significantly increases. For the other limit with $\eta=1$, the spatial overlap starts to increase for $B > 111~$G, a magnetic field larger than $B_0$. The dashed vertical line in Figure~\ref{fig2} represents the critical point for the transition at $B_0$ with $\delta=0$.

\begin{figure}
\includegraphics[width=13.5 cm]{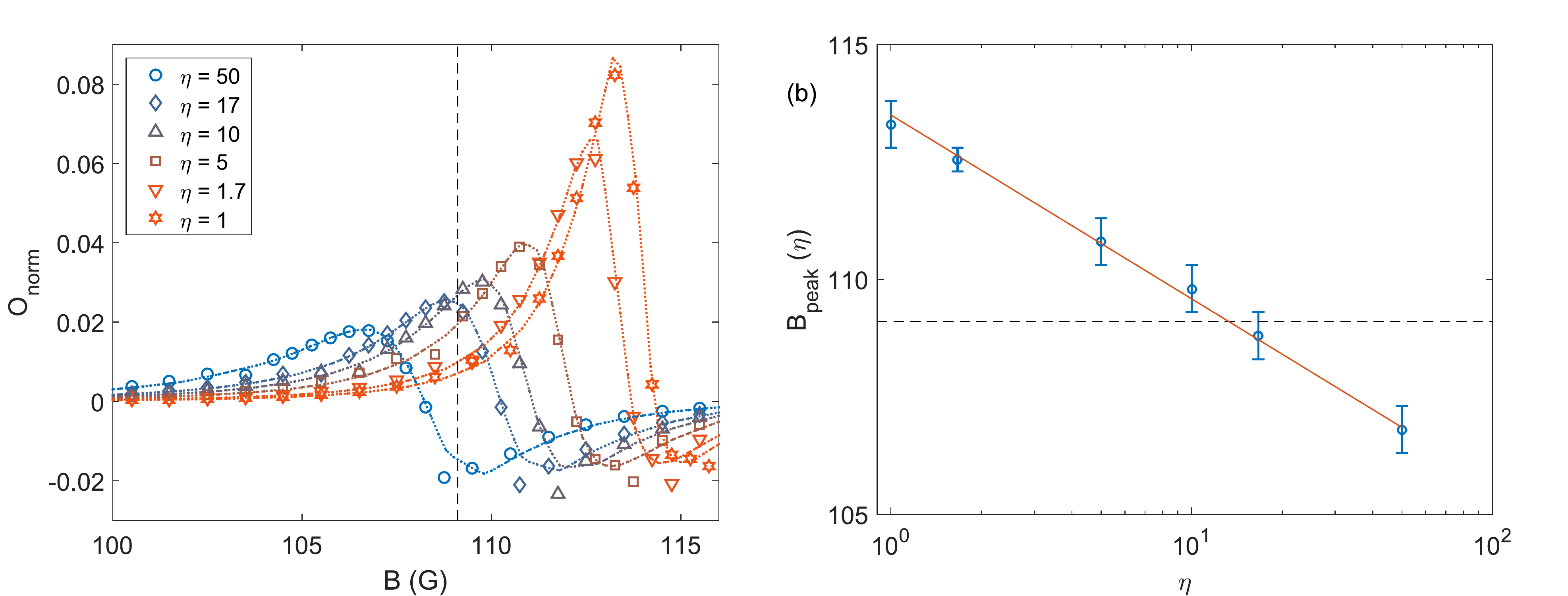}
\caption{\textbf{(a) Numerical second derivative of the normalized overlap, $O_{\text{norm}}$. We identify the peak position of each curve as the magnetic field value, $B_{\text{peak}}$, for which the normalized overlap changes the most indicating the transition from immiscible to miscible. The dotted lines serve only as guide to the eyes. In~(b), we show $B_\text{peak}$ as a function of $\eta$ in a semilog scale which gives a logarithm dependence of $B_\text{peak}$ with the atom number ration. The red solid curve is a fit to the data points (see main text) and  the black dashed line represents $B_0$ with $\delta=0$.}\label{fig4}}
\end{figure}

To define the transition from immiscible to miscible from the normalized overlap we associate a threshold-like behaviour and identify the Feshbach field ($B_\text{peak}$) for which $O_{\text{norm}}$ varies the most. This is done performing the numerical second derivative of the normalized overlap and identifying its maximum value. The second derivatives as a function of the Feshbach field for all $\eta$ are displayed in Figure.~\ref{fig4}. The maximum value of the curves drifts to larger magnetic fields as $\eta$ decreases. In Fig.~\ref{fig4}~(b), we show $B_\text{peak}$ as a function of $\eta$. The almost linear behaviour of the points in the semilog scale suggests a dependence of $B_\text{peak}(\eta)= B_\text{peak}^0-\alpha\ln{\eta}$. We find $B_\text{peak}^0=113.5~$G and $\alpha = 1.70$.

\begin{figure}
\includegraphics[width=10.5 cm]{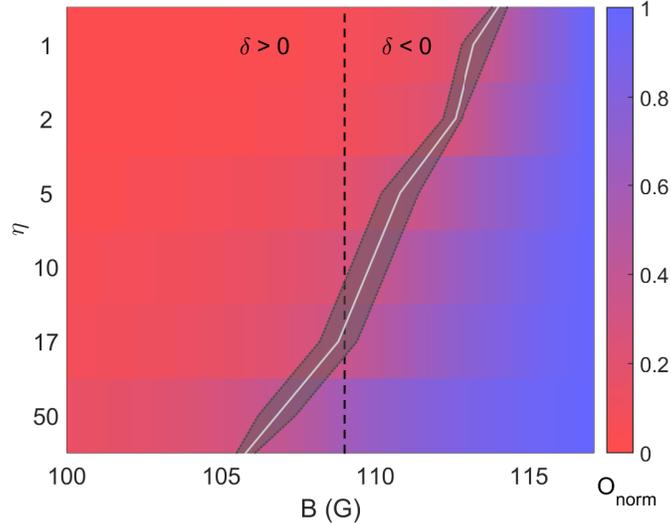}
\caption{\textbf{Phase diagram of the miscible-immiscible phase transition for the $^{23}$Na-$^{39}$K mixture under our experimental conditions. The colormap represents the value of the normalized overlap for each combination of $\eta$ and $B$. The light gray line sets the phase transition point obtained from the second derivative of the normalized overlap. The condition for an homogeneous system is shown by the black dashed line at $B_0=109.1~$G.}.\label{figPhaseDiagram}}
\end{figure}

The miscible-immiscible phase diagram for the $^{23}$Na-$^{39}$K mixture under our experimental conditions is shown in Figure~\ref{figPhaseDiagram}. The colormap represents the value of the normalized overlap for each combination of $\eta$ and $B$ ranging from zero to unit. The transition point for each $\eta$ obtained from Figure~\ref{fig4}~(b) is displayed by the solid light gray curve with the shaded area covering its uncertainty. The black dashed line represents the transition point for the homogeneous case setting $\delta>0$ to the left side of the curve and $\delta<0$ to the right.

\begin{figure}
\includegraphics[width=10.5 cm]{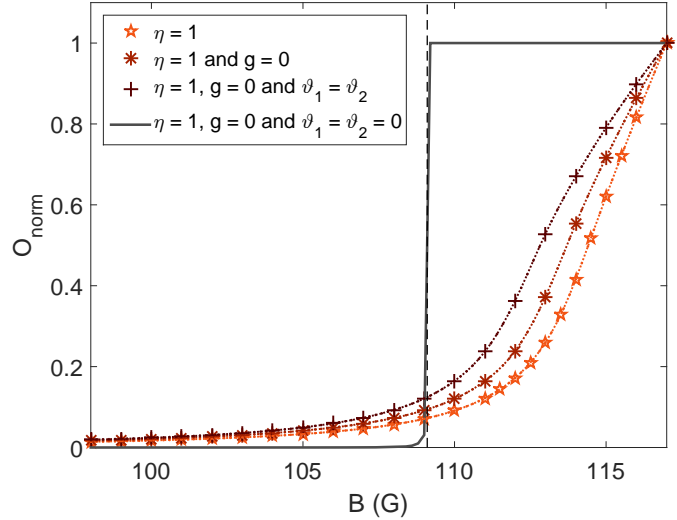}
\caption{Normalized overlap, $O_\text{norm}$, as a function of the Feshbach field, $B$, for $\eta = 1$ under different trapping conditions (see main text).\label{fig3}}
\end{figure}

Differently from earlier works performed with a balanced mixture of two distinct hyperfine states of a single species~\cite{Wen2012PhaseSepAndConfinement,Navarro2009PhaseSep,Bisset2018QuantumSpinFluct2BECs}, in our case for $\eta= 1$ the system is more immiscible and the miscible phase only occurs for $\delta \sim -0.5$. A deeper analysis of the $\eta = 1$ case is presented in Fig.~\ref{fig3}, where we show the normalized overlap for $\eta = 1$ under different trapping conditions: real experimental conditions (star), without gravity (asterisk), considering equal trapping potentials with $\vartheta_1 = \vartheta_2$ (plus sign) and for the homogeneous case (gray solid curve)  obtained setting the external potentials $\vartheta_1(\bold{r})=\vartheta_2(\bold{r})=0$. The dashed vertical line in black represents the point for $\delta=0$ which indeed matches the abrupt transition of $O_\text{norm}$ from $0$ to unity observed in the homogeneous case. The role of gravity and different trapping configurations for each species is clear in the data of Figure~\ref{fig3}: setting $g=0$ and $\vartheta_1=\vartheta_2$ drifts the transition point to smaller magnetic fields approaching $B_0$. However, due to the large difference between the intraspecies scattering lengths for sodium and potassium ($a_\text{Na}=52a_0$ and $a_\text{K}\sim 7.59-8.73a_0$), the system still behaves more immiscible than the homogeneous case.


The identification of the miscibility regime of the Na-K mixture under realistic experimental conditions is important when defining the best parameters for studying different physical phenomena. In studies which the spatial overlap between the components of the mixture is important (i.e. coupled vortex dynamics~\cite{Tsubota2003CoupledVortex,Aftalion2011CoupledVortexLattice,Kuopanportti2012ExoticVortexLattices}, binary quantum turbulence~\cite{Tsubota2010BinaryQT}, coupled superfluidity and excitations~\cite{Ferrari2018SpinSuperfluidity,Shin2020TwoSoundModes}, etc.) it is not always sufficient to have $\delta<0$. The contrary is also true, when the immiscible nature of the system is relevant (i.e. in studies of dynamical instabilities~\cite{Tsubota2006ModInstability,Saito2012RayleighTaylor,Parker2018KelvinHelmholtz,Schmelcher2020StarShapedPatterns}), $\delta>0$ is not always sufficient, specially in the case of large atom number imbalance between the atomic species.






\section{Discussion}\label{Sec:discussion}

We have shown that the miscible-immiscible phase transition in a trapped two-component Bose-Einstein condensation of different atomic species under realistic experimental parameters (considering the effect of gravity and different trapping potentials) suffers strong influence of the atom number ratio $\eta$. In the case of large $\eta$, the system behaves more miscible than the homogeneous case with the transition occurring at $\delta > 0$, while for $\eta=1$, the system is more immiscible with the transition occurring at $\delta<0$. We have defined the miscibility regime of the system by identifying the magnetic field $B_\text{peak}$ for which the normalized spatial overlap between the atomic clouds changes the most. This value was obtained from the magnetic field for which the numerical second derivative of the normalized overlap exhibits a maximum. The behaviour of $B_\text{peak}$ with $\eta$ could be easily associated with a logarithm dependence from the graph of Fig.~\ref{fig4}~(b) making it possible to draw the critical curve in the phase diagram of the miscible-immiscible phase transition for the simulated $^{23}$Na-$^{39}$K quantum mixture (see Fig.~\ref{figPhaseDiagram}). The use of the spatial overlap to identify the miscibility regime of the system could be directly implemented in real experiments by performing high resolution \textit{in situ} images of each atomic species. Further characterizations both on the experimental and theoretical sides could be performed using dynamical properties of the atomic mixture, such as the dipole oscillations proposed in~\cite{Arlt2016PhaseSeparation}, and considering finite temperature effects as realized in recent works~\cite{PhysRevLett.123.075301,PhysRevA.102.063303,roy2021finite}.


\section{Acknowledgements}

The authors thank R. C. Teixeira for sharing a simplified version of the numerical simulations discussed in this manuscript and P. Mazo for providing experimental support in previous stages of the experiment. This work was was funded by S\~ao Paulo Research Foundation (FAPESP) under the grants 2013/07276-1 and 2014/50857-8, and by the National Council for Scientific and Technological Development (CNPq) under the grants 465360/2014-9.

\bibliography{ref_bibitex}

\end{document}